\documentclass[conference]{IEEEtran}
\IEEEoverridecommandlockouts
\usepackage{amsmath,amsfonts}
\usepackage{algorithm,algpseudocode}
\usepackage{array}
\usepackage[caption=false,font=normalsize,labelfont=sf,textfont=sf]{subfig}
\usepackage{textcomp}
\usepackage{stfloats}
\usepackage{url}
\usepackage{verbatim}
\usepackage{graphicx}
\usepackage{cite}
\usepackage{xcolor}
\usepackage[nolist,nohyperlinks]{acronym} 
\usepackage{amsmath,amssymb,amsfonts}
\usepackage{graphicx}
\usepackage{textcomp}
\usepackage{caption}
\usepackage{subcaption}
\usepackage{xspace} 

\usepackage{etoolbox}\AtBeginEnvironment{algorithmic}{\small}

\def\BibTeX{{\rm B\kern-.05em{\sc i\kern-.025em b}\kern-.08em
    T\kern-.1667em\lower.7ex\hbox{E}\kern-.125emX}}

\newcommand{\ten}[1]{\boldsymbol{\mathcal #1}}

\usepackage{tikz,tikz-3dplot}
\usetikzlibrary{angles, calc, positioning, quotes, intersections,fit,decorations.pathreplacing}
\tdplotsetmaincoords{70}{120}

\tikzset{
    cross/.pic = {
    \draw[rotate = 45] (-#1,0) -- (#1,0);
    \draw[rotate = 45] (0,-#1) -- (0, #1);
    }
}

\usepackage{titlesec}

\begin{document}
    % Space between lines
    \renewcommand{\baselinestretch}{1}
    % Space before and after section/subsection/subsubsection
    \titlespacing*{\section}{0pt}{0.2\baselineskip}{0.3\baselineskip}
    \titlespacing*{\subsection}{0pt}{0.2\baselineskip}{0.3\baselineskip}
    \titlespacing*{\subsubsection}{0pt}{0.2\baselineskip}{0.3\baselineskip}
    % Space before and after equation
    \setlength{\abovedisplayskip}{4pt}
    \setlength{\belowdisplayskip}{4pt}
    
    \bstctlcite{IEEE:BSTcontrol}
    \begin{acronym}
        \acro{RIS}{reconfigurable intelligent surface}
        \acro{ALS}{alternating least squares}
        \acro{CE}{channel estimation}
        \acro{RF}{radio-frequency}
        \acro{THz}{Terahertz communication}
        \acro{EVD}{eigenvalue decomposition}
        \acro{CRB}{Cramér-Rao lower bound}
        \acro{CSI}{channel state information}
        \acro{ESPRIT}{estimation of the signal parameters via rotational invariance techniques}
        \acro{BS}{base station}
        \acro{LS}{least squares}
        \acro{SISO}{single-input single-output}
        \acro{MIMO}{multiple-input multiple-output}
        \acro{NMSE}{normalized mean squared error}
        \acro{2G}{Second Generation}
        \acro{3G}{3$^\text{rd}$~Generation}
        \acro{3GPP}{3$^\text{rd}$~Generation Partnership Project}
        \acro{4G}{4$^\text{th}$~Generation}
        \acro{5G}{5$^\text{th}$~Generation}
        \acro{6G}{6$^\text{th}$~generation}
        \acro{E-TALS}{\textit{enhanced} TALS}
        \acro{ULA}{uniform linear array}
        \acro{UT}{user terminal}
        \acro{UTs}{users terminal}
        \acro{LS}{least squares}
        \acro{KRF}{Khatri-Rao factorization}
        \acro{KF}{Kronecker factorization}
        \acro{MU-MIMO}{multi-user multiple-input multiple-output}
        \acro{MU-MISO}{multi-user multiple-input single-output}
        \acro{MU}{multi-user}
        \acro{NTBS}{nested tensor decomposition based sensing}
        \acro{SER}{symbol error rate}
        \acro{SNR}{signal-to-noise ratio}
        \acro{SVD}{singular value decomposition}
        \acro{OFDM}{orthogonal frequency division multiplexing}
        \acro{EoA}{elevation angle of arrival}
        \acro{AoA}{azimuth angle of arrival}
        \acro{EoD}{elevation angle of departure}
        \acro{AoD}{azimuth angle of departure}
        \acro{AWGN}{additive white Gaussian noise} 
        \acro{RCS}{radar cross section}
        \acro{ROT}{Rank-One-Tensor}
        \acro{HOSVD}{higher-order singular value decomposition}
        \acro{LOS}{line-of-sight}
        \acro{NLOS}{non-line-of-sight}
        \acro{GLRT}{generalized likelihood ratio test}
        \acro{CRLB}{Cramér–Rao lower bound}
        \acro{ISAC}{integrated sensing and communications}
        \acro{DFT}{discrete Fourier transform}
        \acro{RMSE}{root mean squared error}
    \end{acronym}
    \title{RIS-Assisted Sensing: A Nested Tensor Decomposition-Based Approach \\
    \thanks{This work was partially supported by the Ericsson Research, Sweden, and Ericsson Innovation Center, Brazil, under UFC.51. The authors acknowledge the partial support of Fundação Cearense de Apoio ao Desenvolvimento Científico e Tecnológico (FUNCAP) under grants FC3-00198-00056.01.00/22 and  ITR-0214-00041.01.00/23, the National Institute of Science and Technology (INCT-Signals) sponsored by Brazil's National Council for Scientific and Technological Development (CNPq) under grant 406517/2022-3, and the Coordenação de Aperfeiçoamento de Pessoal de Nível Superior - Brazil (CAPES) - Finance Code 001. This work is also partially supported by CNPq under grants 312491/2020-4 and 443272/2023-9. G. Fodor was supported by the EU Horizon Europe Program, Grant No: 101139176 6G-MUSICAL.}
    }
    
    \author{\IEEEauthorblockN{Kenneth Benício$^{\ast \dagger}$, Fazal-E-Asim$^{\ast}$, Bruno Sokal$^{\ast}$, André L. F. de Almeida$^{\ast}$, \\ Behrooz Makki$^{\star}$, Gabor Fodor$^{\star}$, and A. Lee Swindlehurst$^{\dagger}$}
    \IEEEauthorblockA{
    \textit{$^\ast$Federal University of Ceara}, Brazil; \textit{$^\star$Ericsson Research}, Sweeden; \textit{$^\dagger$University of California Irvine}, United States  \\
    $^\ast$(kenneth,fazalasim,brunosokal,andre)@gtel.ufc.br; $^\star$(gabor.fodor,behrooz.maki)@ericsson.com; $^\dagger$swindle@uci.edu}
    }
    
    \maketitle
    \thispagestyle{plain}
    \pagestyle{plain}
    \begin{abstract}
        We study a monostatic multiple-input multiple-output sensing scenario assisted by a reconfigurable intelligent surface using tensor signal modeling. We propose a method that exploits the intrinsic multidimensional structure of the received echo signal, allowing us to recast the target sensing problem as a nested tensor-based decomposition problem to jointly estimate the delay, Doppler, and angular information of the target. We derive a two-stage approach based on the alternating least squares algorithm followed by the estimation of the signal parameters via rotational invariance techniques to extract the target parameters. Simulation results show that the proposed tensor-based algorithm yields accurate estimates of the sensing parameters with low complexity.
    \end{abstract}

    \acresetall
    
    \begin{IEEEkeywords}
        Monostatic sensing, reconfigurable intelligent surfaces, parameter estimation, nested tensor-based decomposition.
    \end{IEEEkeywords}
    
    \section{Introduction}
        \IEEEPARstart{R}{econfigurable} intelligent surfaces (RISs) have grown rapidly as a transformative technology and one of the most promising advancements for the next generation of wireless communication systems. \acs{RIS}s are important for mitigating coverage holes in wireless networks and overcoming blockages, especially for higher frequency bands \cite{Rui_Zhang_2021}. \acs{RIS}s also have the potential to improve sensing/tracking tasks performed by the \ac{BS} that do not fall under the conventional communications-only umbrella of the \ac{BS}. Therefore, \acs{RIS}s have the potential to be an integral component in supporting \ac{ISAC} technology \cite{Fan_2022}.
        
        In \cite{zhang2022metaradar}, the authors considered an \acs{RIS} to improve target detection in \ac{MIMO} radar systems by properly adjusting the \acs{RIS} phase shifts and the radar waveform. The authors in \cite{rihan2022spatial} deal with how to deploy an \acs{RIS} to improve the spatial diversity of a radar system by creating an additional virtual \ac{LOS} path between the radar and the desired target. The authors in \cite{song2023intelligent} investigate an \acs{RIS}-assisted \ac{NLOS} sensing scenario and propose both an active and passive beamforming design that minimizes the Cramér–Rao lower bound. The work in \cite{shao2024target} proposes a scenario with a target-mounted \acs{RIS} that aims to enhance the sensing performance for the desired targets while preventing the detection of the target by eavesdroppers. Finally, \cite{kemal2024ris} proposes a parameter estimation method for an \acs{RIS}-assisted monostatic scenario in which a generalized likelihood ratio test is designed. The approach involves a four-dimensional search over the delay-Doppler-azimuth-elevation parameters that uses a repetitive \acs{RIS} phase-shift profile to decompose the four-dimensional search into separate delay-Doppler and azimuth-elevation search problems.  
        
        Our previous work in \cite{10639156} addresses the problem of parameter estimation for a \ac{SISO} sensing system by exploiting the intrinsic multidimensional structure of the received echo signal at the \ac{BS}. In this paper, we generalize the idea to the \ac{MIMO} case. To cope with this more challenging scenario, we propose a \ac{NTBS} algorithm for \acs{RIS}-assisted monostatic sensing \cite{ximenes2015semi}. Our algorithm estimates the desired parameters by exploiting the inherent geometrical structure of the problem and recasting the received echo signal as a fourth-order nested tensor model. The \ac{NTBS} method is a two-stage algorithm based on \ac{ALS}, which initially solves the parameter estimation problem iteratively and, in the final step, employs the well-known \ac{ESPRIT} algorithm \cite{tschudin1999comparison,feng2017comparison} to extract the delay, Doppler, and angle estimates. Our simulation results show the impact of the system parameters on estimation performance, as well as computational complexity. 
        
        \textit{\textbf{Notation}}: Scalars, vectors, matrices, and tensors are respectively represented as $a, \boldsymbol{a}, \boldsymbol{A}$, and $\boldsymbol{\mathcal{A}}$. Also, $\boldsymbol{A}^{*}$, $\boldsymbol{A}^{\text{T}}$, $\boldsymbol{A}^{\text{H}}$, and $\boldsymbol{A}^{\dagger}$ respectively stand for the conjugate, transpose, Hermitian transpose, and pseudoinverse of a matrix $\boldsymbol{A}$. The $j$th column of $\boldsymbol{A} \in \mathbb{C}^{I \times J}$ is denoted by $\boldsymbol{a}_{j} \in \mathbb{C}^{I \times 1}$. The operator D$(\cdot)$ converts a vector into a diagonal matrix, $\text{D}_j(\boldsymbol{B})$ forms a diagonal matrix $R \times R$ out of the $j$th row of $\boldsymbol{B} \in \mathbb{C}^{J \times R}$. The matrix $\boldsymbol{I}_{N}$ denotes an identity matrix of size $N \times N$. The $n$th mode product between a tensor and a matrix is given by $\boldsymbol{\mathcal{A}} \times_{n} \boldsymbol{B} = \boldsymbol{B} \left[\boldsymbol{\mathcal{A}}\right]_{(n)}$. The symbols $\otimes$, $\diamond$, and $\odot$ indicate the Kronecker, Khatri-Rao, and Hadamard products, respectively. Element-wise division is indicated by $\oslash$.
         
    \section{System Model}
        \begin{figure}[!t]
            \centering
            \begin{tikzpicture}[scale=0.975, every node/.style={scale=0.975}]
                \begin{scope}[
                    box1/.style={draw=black, thick, rectangle,rounded corners, minimum height=0.5cm, minimum width=0.5cm}]
                    % The first RIS
                    \node (RIS 1) at (-2,0.75) {$\text{RIS}$};
                    \draw[black,dashed,fill=red!30] (-3.5,-2.5) rectangle (-.5,.5);
                    \node[box1, fill=green!30] (c1) at (-3.,0) {};
                    \node[box1, fill=green!30, right=.125cm of c1] (c2) {};
                    \node[box1, fill=green!30, right=.1265cm of c2] (c3) {};
                    \node[box1, fill=green!30, right=.125cm of c3] (c4) {};
                    \node[box1, fill=green!30, below=.125cm of c4] (c5) {};
                    \node[box1, fill=green!30, left=.125cm of c5] (c6) {};
                    \node[box1, fill=green!30, left=.125cm of c6] (c7) {};
                    \node[box1, fill=green!30, left=.125cm of c7] (c8) {};
                    \node[box1, fill=green!30, below=.125cm of c8] (c9) {};
                    \node[box1, fill=green!30, right=.125cm of c9] (c10) {};
                    \node[box1, fill=green!30, right=.125cm of c10] (c11) {};
                    \node[box1, fill=green!30, right=.125cm of c11] (c12) {};
                    \node[box1, fill=green!30, below=.125cm of c12] (c13) {};
                    \node[box1, fill=green!30, left=.125cm of c13] (c14) {};
                    \node[box1, fill=green!30, left=.125cm of c14] (c15) {};
                    \node[box1, fill=green!30, left=.125cm of c15] (c15) {};
                    
                    %% The dual function BS
                    \draw[black,fill=blue!30] (-7+1.5,-3.5) -- (-6.5+1.5,-3.5) node[above]{$\text{BS}$} -- (-6+1.5,-3.5) -- (-6.5+1.5,-2.4) -- cycle;
                    % The antennas
                    \draw[line width=0.25mm,black] (-7+1.5,-2.4) -- (-6+1.5,-2.4) node[midway,above]{$\cdots$};
                    \draw[line width=0.25mm,black] (-7+1.5,-2.4) -- (-7+1.5,-2.25);
                    \draw[line width=0.25mm,black] (-6+1.5,-2.4) -- (-6+1.5,-2.25);
                    \draw[line width=0.25mm,black] (-7.20+1.5,-2.15) -- (-7+1.5,-2.25) -- (-6.80+1.5,-2.15) -- cycle;
                    \draw[line width=0.25mm,black] (-6.20+1.5,-2.15) -- (-6+1.5,-2.25) -- (-5.80+1.5,-2.15) -- cycle;
    
                    % The cluster of targets
                    \node[circle,draw=black,fill=blue!30,minimum size=12pt] (T1) at (2.70-2,-2-1) {};
                    \node[above] (Target) at (2.70-2,-1.5-1.26) {$\text{Target}$};
    
                    % Drawing the arrows
                    %BS to RIS
                    \draw[line width=0.75mm,black,->] (-6.5+1.5,-1.65) -- (-3.75,0) node[midway, above, rotate=+50]{};
                    \draw[line width=0.75mm,red,<-] (-6.5+1.5,-1.95) -- (-3.75,-0.30) node[midway, above, rotate=+50]{};
                    %RIS to Cluster
                    \draw[line width=0.75mm,black,->] (-0.25,0) -- (2-1.15,-1.65);
                    \draw[line width=0.75mm,red,<-] (-0.25,-0.30) -- (2-1.15,-1.95);
                    % Direct link
                    \draw[line width=0.75mm,dashed,black!75,->] (-6+1.5,-3) -- (-2.25,-3) node[midway, above, rotate=+50]{};
                    \draw (-1.95,-3) pic[rotate = 0] {cross=7pt};
                    % Some labels
                    \node[] () at (-4.65,-0.50) {};
                    \node[] () at (-3.85,-1) {};
                    \node[] () at (+0.5,-0.50) {};
                    \node[] () at (-0.10,-1) {};
                \end{scope}
            \end{tikzpicture}
            \caption{RIS-assisted MIMO Monostatic Sensing}
            \label{fig:system_model}
        \end{figure}
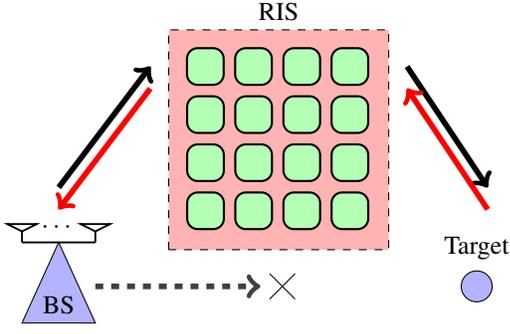
        \begin{figure}[!t]
            \centering
            \begin{tikzpicture}[scale=0.90, every node/.style={scale=0.725}]
                \begin{scope}[
                    box1/.style={draw=black, thick, rectangle, minimum height=0.5cm, minimum width=0.5cm}]
                    % The first block
                    \draw[black,dashed,red,fill=white] (-4.65,-.85) rectangle (-2+0.35,.5);
                    \node[box1, fill=green!30] (c1) at (-4.25,0) {$\boldsymbol{w}_{1}$};
                    \node[box1, fill=green!30, right=.125cm of c1] (c2) {$\boldsymbol{w}_{1}$};
                    \node[right=.125cm of c2] (c3) {$\cdots$};
                    \node[box1, fill=green!30, right=.125cm of c3] (c4) {$\boldsymbol{w}_{1}$};
                    % The second block
                    \draw[black,dashed,red,fill=white] (-1.50,-.85) rectangle (1.15+0.35,.5);
                    \node[box1, fill=green!30] (c1) at (-1.10,0) {$\boldsymbol{w}_{2}$};
                    \node[box1, fill=green!30, right=.125cm of c1] (c2) {$\boldsymbol{w}_{2}$};
                    \node[right=.125cm of c2] (c3) {$\cdots$};
                    \node[box1, fill=green!30, right=.125cm of c3] (c4) {$\boldsymbol{w}_{2}$};
                    % The dots
                    \node[right=.20 of c4,circle,draw=black,fill=black,minimum size=1pt,scale=0.55] (T1) {};
                    \node[right=.1 of T1,circle,draw=black,fill=black,minimum size=1pt,scale=0.55] (T2) {};
                    \node[right=.1 of T2,circle,draw=black,fill=black,minimum size=1pt,scale=0.55] (T3) {};
                    % The third block
                    \draw[black,dashed,red,fill=white] (2.35,-.85) rectangle (5.35,.5);
                    \node[box1, fill=green!30] (c1) at (2.75,0) {$\boldsymbol{w}_{K}$};
                    \node[box1, fill=green!30, right=.125cm of c1] (c2) {$\boldsymbol{w}_{K}$};
                    \node[right=.125cm of c2] (c3) {$\cdots$};
                    \node[box1, fill=green!30, right=.125cm of c3] (c4) {$\boldsymbol{w}_{K}$};
                    % The arrows
                    \draw[line width=0.15mm,black,<->] (-4.65,-1) -- (-1.65,-1) node[midway, below, rotate=+0]{\large Block size ($Q M$)};
                    \draw[line width=0.15mm,black,<->] (-4.65,0.75) -- (5.35,0.75) node[midway, above, rotate=+0]{\large Total of $K Q M$ OFDM symbols};
                    \draw[line width=0.15mm,black,<->] (-4.45,-0.5) -- (-2.25+0.35,-0.5) node[midway, below, rotate=+0]{$1$st block};
                    \draw[line width=0.15mm,black,<->] (-1.3,-0.5) -- (1+0.35,-0.5) node[midway, below, rotate=+0]{$2$nd block};
                    \draw[line width=0.15mm,black,<->] (2.5,-0.5) -- (4.85+0.35,-0.5) node[midway, below, rotate=+0]{$K$th block};
                \end{scope}
            \end{tikzpicture}
            \caption{Time-domain protocol of the proposed RIS design.}
            \label{fig:transmission_protocol}
        \end{figure}
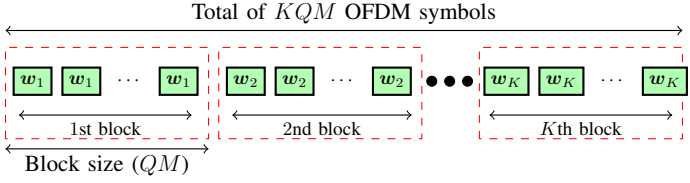
        
        Consider a downlink monostatic \acs{RIS} scenario, as shown in Figure \ref{fig:system_model}. The \ac{BS} has multiple antennas and is assisted by an \acs{RIS} with $N = N_yN_z$ passive reflecting elements. The direct \ac{LOS} between the BS and the target is blocked. The target reflection is only accessible via a virtual \ac{LOS} path via the \acs{RIS}. Furthermore, the \ac{BS} transmits an \ac{OFDM} waveform towards the target and listens to the echo backscattered from the desired target via the \acs{RIS}. The \ac{BS} is equipped with an $L$-element uniformly-spaced antenna array and transmits an \ac{OFDM} pulse with $Q$ subcarriers and $M$ symbols, structured as shown in Figure \ref{fig:transmission_protocol}. The complex pilot on the $q$th subcarrier for the $m$th symbol is denoted as $\boldsymbol{x}_{q,m} \in \mathbb{C}^{L \times 1}$. The \ac{OFDM} echo signal received via the \acs{RIS}-assisted path is given by \label{SEC:System_Model}
        \begin{equation}
            \begin{aligned}
                \hspace{-0.15cm}\boldsymbol{y}_{q,m,k} \hspace{-0.10cm}= &\alpha \underbrace{\boldsymbol{a}(\eta) \boldsymbol{b}^{\text{T}} (\boldsymbol{\phi}) \text{D}(\boldsymbol{w}_{k}) \boldsymbol{p}(\boldsymbol{\theta})}_{\text{Target-\acs{RIS}-\ac{BS} path}} \underbrace{\boldsymbol{p}^{\text{T}}(\boldsymbol{\theta}) \text{D}(\boldsymbol{w}_{k}) \boldsymbol{b}(\boldsymbol{\phi}) \boldsymbol{a}^{\text{T}} (\eta)}_{\text{\ac{BS}-\acs{RIS}-Target path}} \\ &\times \boldsymbol{x}_{q,m} \left[\boldsymbol{c}(\tau)\right]_{q} \left[\boldsymbol{d}(\nu)\right]_{m} + \boldsymbol{z}_{q,m} \in \mathbb{C}^{L \times 1},  \label{eq:received_echo_1}
            \end{aligned}
        \end{equation}
        in the frequency domain, where $\boldsymbol{a}(\eta) \in \mathbb{C}^{L \times 1}$ is the steering vector response of the uniform linear array at the \ac{BS} where $\eta = \pi \text{cos}(\kappa)$ is the spatial frequency and $\kappa$ the \ac{AoD} from the \ac{BS} to the \acs{RIS}. Assuming half wavelength element spacing, the transmit steering vector of the \ac{BS} is given by
        \begin{align}
            \boldsymbol{a}(\eta) = \left[1, e^{-j\eta} ,\dots, e^{-j(L-1)\eta}\right]^T \in \mathbb{C}^{L \times 1}.
        \end{align}
        
        The steering vector at the \acs{RIS} corresponding to the \ac{BS}-\acs{RIS} channel is represented by $\boldsymbol{b}(\boldsymbol{\phi}) \in \mathbb{C}^{N \times 1}$ where $\boldsymbol{\phi} = \left[ \mu_\text{A}, \psi_\text{A}\right]^T$ and $\mu_{\text{A}}, \psi_{\text{A}}$ are the horizontal and vertical spatial frequencies. Assuming the RIS lies in the $y$-$z$ plane, the horizontal and vertical spatial frequencies can be defined as $\mu_{\text{A}} = \pi\sin\theta_{\text{A}}\sin\phi_{\text{A}}$ and $\psi_{\text{A}} = \pi\cos\theta_{\text{A}}$, where $\theta_{\text{A}}$ is the \ac{EoA} and $\phi_{\text{A}}$ the \ac{AoA} of the transmitted \ac{OFDM} pulse towards the \acs{RIS} from the \ac{BS} \cite{Asim_2021}. The steering vector $\boldsymbol{b}(\boldsymbol{\phi})$ can be expressed as the Kronecker product between the horizontal and vertical components. Assuming half-wavelength element spacing, we have
        \begin{align}
            \boldsymbol{b}(\boldsymbol{\phi}) = \boldsymbol{b}(\mu_{\text{A}},\psi_{\text{A}}) = \boldsymbol{b}_y(\mu_{\text{A}}) \otimes \boldsymbol{b}_z(\psi_{\text{A}}) \in \mathbb{C}^{N \times 1}, \label{ch_3_kron_channel}
        \end{align}
        where
        \begin{align}
            \boldsymbol{b}_y(\mu_{\text{A}}) = \left[1, e^{-j\mu_{\text{A}}} ,\dots, e^{-j(N_y-1)\mu_{\text{A}}}\right]^T \in \mathbb{C}^{N_y \times 1},
        \end{align}
        and
        \begin{align}
            \boldsymbol{b}_z(\psi_{\text{A}}) = \left[1, e^{-j\psi_{\text{A}}} ,\dots, e^{-j(N_z-1)\psi_{\text{A}}}\right]^T \in \mathbb{C}^{N_z \times 1}.
        \end{align}     
        
        Similarly, $\boldsymbol{p}(\boldsymbol{\theta}) \in \mathbb{C}^{N \times 1}$ is the steering vector corresponding to the channel between the target and the \acs{RIS} and can also be represented as the Kronecker product between the horizontal and vertical components as
        \begin{align}
            \boldsymbol{p}(\boldsymbol{\theta}) = \boldsymbol{p}(\mu_{\text{D}},\psi_{\text{D}}) = \boldsymbol{p}_y(\mu_{\text{D}}) \otimes \boldsymbol{p}_z(\psi_{\text{D}}) \in \mathbb{C}^{N \times 1}, \label{ch_4_kron_channel}
        \end{align}
        where $\boldsymbol{\theta} = \left[ \mu_\text{D}, \psi_\text{D}\right]^T$ contains horizontal and vertical spatial frequencies of departure given as $\mu_{\text{D}} = \pi\sin\theta_{\text{D}}\sin\phi_{\text{D}}$, and $\psi_{\text{D}} = \pi\cos\theta_{\text{D}}$,
        with $\phi_\text{D}$ and $\theta_{\text{D}}$ representing the \ac{AoD} and \ac{EoD}, respectively. The horizontal and vertical steering vectors of departure can be represented as $\boldsymbol{p}_y(\mu_{\text{D}}) = \left[1, e^{-j\mu_{\text{D}}} ,\dots, e^{-j(N_y-1)\mu_{\text{D}}}\right]^T \in \mathbb{C}^{N_y \times 1}$ and $\boldsymbol{p}_z(\psi_{\text{D}}) = \left[1, e^{-j\psi_{\text{D}}},\dots, e^{-j(N_z-1)\psi_{\text{D}}}\right]^T \in \mathbb{C}^{N_z \times 1}$. 
        
        The \acs{RIS}  phase-shift vector $\boldsymbol{w}_{k} \in \mathbb{C}^{N \times 1}$ for the $k$th transmission block is defined as $\boldsymbol{w}_{k} = \left[e^{j \zeta_{1,k}} , \cdots, e^{j \zeta_{N,k}} \right]^{\text{T}} \in \mathbb{C}^{N \times 1}$, where $\zeta_{n,k}$ is the phase-shift of the $n$th \acs{RIS} element in the $k$th block. The frequency-domain steering vector due to the time delay $\tau$ is $\boldsymbol{c}(\tau) = \left[1, \cdots, e^{-j 2 \pi (Q - 1) \Delta f \tau}  \right]^{\text{T}} \in \mathbb{C}^{Q \times 1}$, whose $q$th element $\left[\boldsymbol{c}(\tau)\right]_{q}$ is linked to the $q$-th \ac{OFDM} subcarrier. Furthermore, $\boldsymbol{d}(\nu) = \left[1, \cdots, e^{j 2 \pi (M - 1) T_{s} \nu}\right]^{\text{T}} \in \mathbb{C}^{M \times 1}$ is the time-domain steering vector corresponding to the Doppler shift $\nu$, and it is linked to the $m$th \ac{OFDM} symbol by the $m$th element $\left[\boldsymbol{d}(\nu)\right]_{m} \in \mathbb{C}$. Finally, $\boldsymbol{z}_{q,m} \in \mathbb{C}^{L \times 1}$ is \ac{AWGN} at the $q$th subcarrier of the $m$th symbol. The overall magnitude of the complex gain, $\alpha \in \mathbb{C}$, can be written as \cite{kemal2024ris}
        \begin{align}
            |\alpha| = \sqrt{\frac{P_{t} G^{2}_{1} G^{2}_{2} F^{2}_{1}(\phi_{\text{el}}) F^{2}_{2}(\theta_{\text{el}}) d^{2}_{x} d^{2}_{y} \lambda^{2} \sigma_{\text{RCS}}}{(4\pi)^{5} d^{4}_{1}d^{4}_{2}}},
        \end{align}
        \noindent where $P_{t}$ denotes the transmit power, $G_{1}$ denotes \ac{BS} transmit antenna gain, $G_{2}$ is the \ac{BS} receive antenna gain, $F^{2}_{1}(\phi_{\text{A}})$ is the normalized \acs{RIS} power radiation pattern at the \ac{AoA} $\phi_{\text{D}}$, $F^{2}_{2}(\theta_{\text{A}})$ is the normalized \acs{RIS} power radiation pattern at the \ac{AoD} $\theta_{\text{D}}$, $d_{x}$ and $d_{y}$ denote the \acs{RIS} spacing along the horizontal and vertical domains, respectively, $\lambda$ is the carrier wavelength, $\sigma_{\text{RCS}}$ is the target radar cross section, and $d_{1}$ and $d_{2}$ denote the \ac{BS}-\acs{RIS} and \acs{RIS}-target distances, respectively. 
        
        Writing Equation (\ref{eq:received_echo_1}) in compact form leads to
        \begin{align}
            \begin{split}
                \boldsymbol{y}_{q,m,k} = \alpha \boldsymbol{H} \text{D}(\boldsymbol{w}_{k}) \boldsymbol{p}(\boldsymbol{\theta})\boldsymbol{p}^{\text{T}}(\boldsymbol{\theta}) \text{D}(\boldsymbol{w}_{k}) \boldsymbol{H}^{\text{T}} \boldsymbol{x}_{q,m} \\ \times \left[\boldsymbol{c}(\tau)\right]_{q} \left[\boldsymbol{d}(\nu)\right]_{m} + \boldsymbol{z}_{q,m} \in \mathbb{C}^{L \times 1}, 
            \end{split} \label{eq:received_echo_2}
        \end{align}
        where $\boldsymbol{H} = \boldsymbol{a}(\eta) \boldsymbol{b}^{\text{T}}(\boldsymbol{\phi}) \in \mathbb{C}^{L \times N}$ is the compact form of the \ac{BS}-\acs{RIS} channel. By stacking columnwise the data from all the $Q$ subcarriers and $M$ OFDM symbols, we obtain the following noiseless signal
        \begin{align}
            \boldsymbol{Y}_{k} &\hspace{-0.075cm}=\hspace{-0.1cm} \boldsymbol{H} \text{D}(\boldsymbol{w}_{k}) \boldsymbol{p}(\boldsymbol{\theta}) \boldsymbol{p}^{\text{T}}(\boldsymbol{\theta}) \text{D}(\boldsymbol{w}_{k}) \boldsymbol{H}^{\text{T}} \boldsymbol{X} \text{D}\left(\boldsymbol{c}(\tau) \hspace{-0.075cm}\otimes\hspace{-0.075cm} \boldsymbol{d}(\nu)\right),
        \end{align}
        where $\boldsymbol{X} = \left[\boldsymbol{x}_{1,1}, \cdots, \boldsymbol{x}_{Q,1}, \boldsymbol{x}_{1,2}, \cdots, \boldsymbol{x}_{Q,M}\right]\in \mathbb{C}^{L \times M Q}$. Defining $\boldsymbol{P}(\theta) = \boldsymbol{p}(\theta) \boldsymbol{p}^{\text{T}}(\theta) \in \mathbb{C}^{N \times N}$ and applying the properties $\text{vec}(\boldsymbol{A} \boldsymbol{B} \boldsymbol{C}) = (\boldsymbol{C}^{\text{T}} \otimes \boldsymbol{A}) \text{vec} (\boldsymbol{B})$ and $\text{vec}(\boldsymbol{A} \text{D}(\boldsymbol{b}) \boldsymbol{C}) = (\boldsymbol{C}^{\text{T}} \diamond \boldsymbol{A}) \boldsymbol{b}$ leads to
        \begin{align}
            \begin{split}
                \notag \boldsymbol{y}_{k} &\hspace{-0.05cm}=\hspace{-0.05cm} \left[\left(\boldsymbol{H}^{\text{T}} \boldsymbol{X} \text{D}\left[\boldsymbol{c}(\tau) \otimes \boldsymbol{d}(\nu)\right]\right)^{\text{T}}\hspace{-0.15cm}\otimes\hspace{-0.05cm} \boldsymbol{H}\right] \hspace{-0.10cm} \text{D}(\boldsymbol{w}_{k} \hspace{-0.10cm}\otimes\hspace{-0.05cm} \boldsymbol{w}_{k}) \text{vec}(\boldsymbol{P}(\boldsymbol{\theta})),
            \end{split} \\
            \begin{split}
                \notag &\hspace{-0.05cm}=\hspace{-0.05cm} \left\{\left(\text{vec}(\boldsymbol{P}(\theta))\right)^{\text{T}} \diamond \left[\left(\boldsymbol{H}^{\text{T}} \boldsymbol{X} \text{D}\left[\boldsymbol{c}(\tau) \otimes \boldsymbol{d}(\nu)\right]\right)^{\text{T}}\otimes \boldsymbol{H}\right] \right\} \\ &\left(\boldsymbol{w}_{k} \otimes \boldsymbol{w}_{k} \right) \in \mathbb{C}^{L M Q\times 1},
             \end{split}
        \end{align}
        where $\boldsymbol{y}_{k} = \text{vec}(\boldsymbol{Y}_{k})$. Using the property $\boldsymbol{a}^{\text{T}} \diamond \boldsymbol{B} = \boldsymbol{B} \text{D}(\boldsymbol{a})$ yields
        \begin{align}
            \begin{split}
                \notag \boldsymbol{y}_{k} &= \left[\left(\boldsymbol{H}^{\text{T}} \boldsymbol{X} \text{D}\left[\boldsymbol{c}(\tau) \otimes \boldsymbol{d}(\nu)\right]\right)^{\text{T}}\otimes \boldsymbol{H}\right] \text{D}\left[(\text{vec}(\boldsymbol{P}(\theta)))\right] 
                \\ &\left(\boldsymbol{w}_{k} \otimes \boldsymbol{w}_{k} \right).
            \end{split}
        \end{align}
        Collecting the data from all $K$ symbol periods, we obtain
        \begin{align}
            \notag \boldsymbol{Y} &= \left[\boldsymbol{y}_{1}, \cdots, \boldsymbol{y}_{K}\right] \in \mathbb{C}^{L M Q \times K}, \\
            % \begin{split}
            %    \notag \boldsymbol{Y} &= \left[\left(\boldsymbol{H}^{\text{T}} \boldsymbol{X} \text{D}\left[\boldsymbol{c}(\tau) \otimes \boldsymbol{d}(\nu)\right]\right)^{\text{T}}\otimes \boldsymbol{H}\right] \text{D}\left[(\text{vec}(\boldsymbol{P}(\theta)))\right] \\ &\left[\left(\boldsymbol{w}_{1} \otimes \boldsymbol{w}_{1}\right), \cdots, \left(\boldsymbol{w}_{K} \otimes \boldsymbol{w}_{K}\right)\right],
            % \end{split} \\
            \begin{split} &= \left[\left(\boldsymbol{H}^{\text{T}} \boldsymbol{X} \text{D}\left[\boldsymbol{c}(\tau) \hspace{-0.05cm}\otimes\hspace{-0.05cm} \boldsymbol{d}(\nu)\right]\right)^{\text{T}} \hspace{-0.05cm}\otimes\hspace{-0.05cm} \boldsymbol{H}\right] \hspace{-0.05cm}\text{D}\left[(\text{vec}(\boldsymbol{P}(\theta)))\right] \\ &\left(\boldsymbol{W} \hspace{-0.05cm}\diamond\hspace{-0.05cm} \boldsymbol{W}\right).\end{split} \label{eq:received_echo_unfold}
        \end{align}
    
    \section{Nested Tensor-Based Sensing}
       Note that Equation (\ref{eq:received_echo_unfold}) corresponds to the 3-mode unfolding of a tensor $\boldsymbol{\mathcal{Y}} \in \mathbb{C}^{L \times M Q \times K}$, which follows a Tucker-3 decomposition \cite{comon2009tensor, de2021channel} represented as
        \begin{align}
            \boldsymbol{\mathcal{Y}} &= \boldsymbol{\mathcal{P}} \hspace{-0.05cm}\times_1\hspace{-0.05cm} \boldsymbol{H} \hspace{-0.05cm}\times_2\hspace{-0.05cm} \boldsymbol{F}(\tau, \nu) \hspace{-0.05cm}\times_3\hspace{-0.05cm} \left(\boldsymbol{W} \hspace{-0.05cm}\diamond\hspace{-0.05cm} \boldsymbol{W}\right)^{\text{T}} \label{eq:3rd_order_tucker},
        \end{align}
        where $\boldsymbol{F}(\tau, \nu) = \left(\boldsymbol{H}^{\text{T}} \boldsymbol{X} \text{D}\left[\boldsymbol{c}(\tau) \otimes \boldsymbol{d}(\nu)\right]\right)^{\text{T}} \in \mathbb{C}^{M Q \times N}$ and its matrix unfoldings are written as:
        \begin{align*}
            \left[\boldsymbol{\mathcal{Y}}\right]_{(1)} &= \boldsymbol{H} \left[\boldsymbol{\mathcal{P}}\right]_{(1)} \left[\left(\boldsymbol{W} \diamond \boldsymbol{W}\right)^{\text{T}} \otimes \boldsymbol{F}(\tau, \nu)\right]^{\text{T}} \in \mathbb{C}^{L \times M Q K}, \\
            \left[\boldsymbol{\mathcal{Y}}\right]_{(2)} &= \boldsymbol{F}(\tau, \nu) \left[\boldsymbol{\mathcal{P}}\right]_{(2)} \left[\left(\boldsymbol{W} \diamond \boldsymbol{W}\right)^{\text{T}} \otimes \boldsymbol{H}\right]^{\text{T}} \in \mathbb{C}^{M Q \times L K}, \\ 
            \left[\boldsymbol{\mathcal{Y}}\right]_{(3)} &= \left(\boldsymbol{W} \diamond \boldsymbol{W}\right)^{\text{T}} \left[\boldsymbol{\mathcal{P}}\right]_{(3)} \left[\boldsymbol{F}(\tau, \nu) \otimes \boldsymbol{H}\right]^{\text{T}} \in \mathbb{C}^{K \times L M Q}.
        \end{align*}
        Applying property $\text{vec}(\boldsymbol{A} \text{D}(\boldsymbol{b}) \boldsymbol{C}) = (\boldsymbol{C}^{\text{T}} \diamond \boldsymbol{A}) \boldsymbol{b}$ in Equation (\ref{eq:3rd_order_tucker}), we can write the received echo signal as
        \begin{align*}
            \text{vec}(\left[\boldsymbol{\mathcal{Y}}\right]_{(3)}) &= \left[\left[\boldsymbol{F}(\tau, \nu) \otimes \boldsymbol{H}\right] \diamond \left(\boldsymbol{W} \diamond \boldsymbol{W}\right)^{\text{T}}\right] \text{vec}(\left[\boldsymbol{\mathcal{P}}\right]_{(3)}).
        \end{align*}
        
        \subsection{ALS-Based Estimation}
            In this section, we develop the proposed \ac{NTBS} parameter estimation algorithm. Our algorithm comprises two stages, each based on the \ac{ALS} estimation scheme \cite{comon2009tensor, benicio2023tensor, benicio2023tensor_wcl}. In the first stage, we estimate the tensor constructed from the received echo signal matrix in (\ref{eq:3rd_order_tucker}). We can exploit the intrinsic multidimensionality of $\boldsymbol{F}(\nu,\tau)$, which contains the Doppler and delay information from the target. In the second stage, we build a $3$rd order Tucker model to estimate the Doppler and delay steering vectors. After the second stage, we employ the \ac{ESPRIT} algorithm to estimate the target parameters. The proposed method is summarized in Algorithm \ref{alg:proposed}. 
            
            \textbf{First Stage:} Since the RIS phase-shift matrix $\boldsymbol{W}$ is known at the BS, the tensor in Equation (\ref{eq:3rd_order_tucker}) can be estimated by defining the following tensor fitting problem
                \begin{equation}
                    \hspace{-0.3cm}\left\{\hat{\boldsymbol{H}}, \hat{\boldsymbol{F}}(\tau, \nu), \hat{\boldsymbol{\mathcal{P}}}\right\} \hspace{-0.1cm}=\hspace{-0.2cm} \underset{\boldsymbol{H}, \boldsymbol{F}(\tau, \nu), \boldsymbol{\mathcal{P}}}{\text{arg min}} \hspace{-0.075cm} \left| \hspace{-0.05cm} \left| \begin{aligned}\boldsymbol{\mathcal{Y}} - \boldsymbol{\mathcal{P}} \times_{1} \boldsymbol{H} \times_{2}& \\ \boldsymbol{F}(\tau, \nu) \times_{3} \left(\boldsymbol{W} \diamond \boldsymbol{W}\right)^{\text{T}}&\end{aligned} \right| \hspace{-0.05cm} \right|^{2}_{\text{F}}, \hspace{-0.1cm} \label{eq:tensor_fit_problem_1}
                \end{equation}
                which is solved by means of the \ac{ALS} algorithm and yields estimates of $\boldsymbol{H}$ and $\boldsymbol{F}(\tau, \nu)$ up to scaling ambiguities. The solution estimates the target signatures in an alternating way by iteratively solving the following optimizations:
                \begin{equation}
                     \hspace{-0.2cm}\hat{\boldsymbol{H}} \hspace{-0.1cm}=\hspace{-0.1cm} \underset{\boldsymbol{H}}{\text{arg min}} \left|\left|\begin{aligned}\left[\boldsymbol{\mathcal{Y}}\right]_{(1)} -& \\ \boldsymbol{H} \left[\boldsymbol{\mathcal{P}}\right]_{(1)} \left[\left(\boldsymbol{W} \diamond \boldsymbol{W}\right)^{\text{T}} \otimes \boldsymbol{F}(\tau, \nu)\right]^{\text{T}}&\end{aligned}\right|\right|^{2}_{\text{F}}, \label{eq:ls_1}
                \end{equation}
                \vspace{-0.2cm}
                \begin{equation}
                    \hspace{-0.3cm}\hat{\boldsymbol{F}}(\tau, \nu) \hspace{-0.1cm}=\hspace{-0.1cm} \underset{\boldsymbol{F}(\tau, \nu)}{\text{arg min}} \hspace{-0.075cm} \left|\hspace{-0.05cm}\left|\begin{aligned}\left[\boldsymbol{\mathcal{Y}}\right]_{(2)} -& \\ \boldsymbol{F}(\tau, \nu) \left[\boldsymbol{\mathcal{P}}\right]_{(2)} \left[\left(\boldsymbol{W} \diamond \boldsymbol{W}\right)^{\text{T}} \otimes \boldsymbol{H}\right]^{\text{T}}&\end{aligned}\right|\hspace{-0.05cm}\right|^{2}_{\text{F}}, \label{eq:ls_2}
                \end{equation}
                \vspace{-0.2cm}
                \begin{equation}
                    \hspace{-0.2cm}\text{vec}(\hat{\left[\boldsymbol{\mathcal{P}}\right]}_{(3)}) \hspace{-0.1cm}=\hspace{-0.1cm} \underset{\text{vec}(\left[\boldsymbol{\mathcal{P}}\right]_{(3)})}{\text{arg min}} \hspace{-0.075cm} \left|\hspace{-0.05cm}\left|\begin{aligned}\text{vec}(\left[\boldsymbol{\mathcal{Y}}\right]_{(3)}) - [[\boldsymbol{F}(\tau, \nu) \otimes & \\ \boldsymbol{H}] \diamond \left(\boldsymbol{W} \diamond \boldsymbol{W}\right)^{\text{T}}] \text{vec}(\left[\boldsymbol{\mathcal{P}}\right]_{(3)})&\end{aligned} \right|\hspace{-0.05cm}\right|^{2}_{\text{F}} \hspace{-0.075cm}, \label{eq:ls_3}
                \end{equation}
                \vspace{-0.2cm}
                whose solutions are respectively given by
                \begin{align}
                    \hat{\boldsymbol{H}} &\hspace{-0.05cm}=\hspace{-0.05cm} \left[\boldsymbol{\mathcal{Y}}\right]_{(1)} \left[\hspace{-0.05cm}\left[\boldsymbol{\mathcal{P}}\right]_{(1)}\hspace{-0.05cm} \left[\left(\boldsymbol{W} \hspace{-0.05cm}\diamond\hspace{-0.05cm} \boldsymbol{W}\right)^{\text{T}} \hspace{-0.05cm}\otimes\hspace{-0.05cm} \boldsymbol{F}(\tau, \nu)\right]^{\text{T}}\hspace{-0.05cm}\right]^{\dagger}, \label{eq:1st_stage_pseudoinverse_1} \\
                    \hat{\boldsymbol{F}}(\tau, \nu) &= \left[\boldsymbol{\mathcal{Y}}\right]_{(2)} \left[\left[\boldsymbol{\mathcal{P}}\right]_{(2)} \left[\left(\boldsymbol{W} \diamond \boldsymbol{W}\right)^{\text{T}} \otimes \boldsymbol{H}\right]^{\text{T}}\right]^{\dagger}, \label{eq:1st_stage_pseudoinverse_2} \\
                    \text{vec}(\hat{\left[\boldsymbol{\mathcal{P}}\right]}_{(3)}) &\hspace{-0.1cm}=\hspace{-0.1cm} \left[\left[\boldsymbol{F}(\tau, \nu) \hspace{-0.05cm}\otimes\hspace{-0.05cm} \boldsymbol{H}\right] \hspace{-0.05cm}\diamond\hspace{-0.05cm} \left(\boldsymbol{W} \hspace{-0.05cm}\diamond\hspace{-0.05cm} \boldsymbol{W}\right)^{\text{T}}\right]^{\dagger} \hspace{-0.1cm} \text{vec}(\left[\boldsymbol{\mathcal{Y}}\right]_{(3)}). \label{eq:1st_stage_pseudoinverse_3}
                \end{align}
                
                In the second stage, we exploit the intrinsic multidimensional structure of $\hat{\boldsymbol{F}}(\tau,\nu)$ by reshaping it as a $3$rd order Tucker tensor, and we estimate the steering vectors corresponding to the target delay and Doppler signatures. 
                \begin{algorithm}[!t]
                    \vspace{-0.1cm}
                    \caption{\ac{NTBS}} 
                    \label{alg:proposed}
                    \begin{algorithmic}[1]
                        \Require{Tensor $\boldsymbol{\mathcal{Y}}$, $\boldsymbol{W}$, $\boldsymbol{X}$, maximum number of iterations $i_{\text{max}}$, and convergence threshold $\delta$.}
                        \textbf{First stage}
                        \State{Randomly initialize $\hat{\boldsymbol{H}}$, $\hat{\boldsymbol{F}}(\tau, \nu)$, and $\hat{\boldsymbol{\mathcal{P}}}$ at iteration $i = 0$.}
                        \While{$||e(i) - e(i-1)|| \geq \delta$ and $i < i_{\text{max}}$}
                            \State{Find the \ac{LS} estimate of $\boldsymbol{H}$ as}
                            \vspace{-0.225cm}
                            \begin{align*}
                                \hat{\boldsymbol{H}} &= \left[\boldsymbol{\mathcal{Y}}\right]_{(1)} \left[\hat{\left[\boldsymbol{\mathcal{P}}\right]}_{(1)} \left[\left(\boldsymbol{W} \diamond \boldsymbol{W}\right)^{\text{T}} \otimes \hat{\boldsymbol{F}}(\tau, \nu)\right]^{\text{T}}\right]^{\dagger}
                            \end{align*}
                            \vspace{-0.225cm}
                            \State{Find the \ac{LS} estimate of $\boldsymbol{F}(\tau, \nu)$ as}
                            \vspace{-0.225cm}
                            \begin{align*}
                                \hat{\boldsymbol{F}}(\tau, \nu) &= \left[\boldsymbol{\mathcal{Y}}\right]_{(2)} \left[\hat{\left[\boldsymbol{\mathcal{P}}\right]}_{(2)} \left[\left(\boldsymbol{W} \diamond \boldsymbol{W}\right)^{\text{T}} \otimes \hat{\boldsymbol{H}}\right]^{\text{T}}\right]^{\dagger}
                            \end{align*}
                            \vspace{-0.225cm}
                            \State{Find the \ac{LS} estimate of $\boldsymbol{\mathcal{P}}$ as}
                            \vspace{-0.15cm}
                            \begin{align*}
                                \text{vec}(\hat{\left[\boldsymbol{\mathcal{P}}\right]}_{(3)}) \hspace{-0.1cm}=\hspace{-0.1cm} \left[\left[\hat{\boldsymbol{F}}(\tau, \nu) \hspace{-0.05cm}\otimes\hspace{-0.05cm} \hat{\boldsymbol{H}}\right] \hspace{-0.05cm}\diamond\hspace{-0.05cm} \left(\boldsymbol{W} \hspace{-0.05cm}\diamond\hspace{-0.05cm} \boldsymbol{W}\right)^{\text{T}}\right]^{\dagger} \hspace{-0.1cm} \text{vec}(\left[\boldsymbol{\mathcal{Y}}\right]_{(3)})
                            \end{align*}
                            \State{Update $e(i) = ||\hat{\mathcal{Y}} - \hat{\mathcal{Y}}(i)||^{2}_{\text{F}}$}
                        \EndWhile
                        \State{\textbf{return} $\hat{\boldsymbol{H}}$, $\hat{\boldsymbol{F}}(\tau, \nu)$, and $\hat{\boldsymbol{\mathcal{P}}}$.}
                        \newline
                        \textbf{Second stage}
                        \State{Build tensor $\hat{\boldsymbol{\mathcal{F}}}$ from $\hat{\boldsymbol{F}}(\tau, \nu)$.}
                        \State{Randomly initialize $\hat{\boldsymbol{d}}(\nu)$, and $\hat{\boldsymbol{c}}(\tau)$ at iteration $i = 0$.}
                        \While{$||e(i) - e(i-1)|| \geq \delta$ and $i < i_{\text{max}}$}
                            \State{Find the \ac{LS} estimate of $\boldsymbol{d}(\nu)$ as}
                            \vspace{-0.225cm}
                            \begin{align*}
                                \hat{\boldsymbol{d}}(\nu) &\hspace{-0.05cm}=\hspace{-0.05cm} \left[\left(\left[\boldsymbol{\mathcal{X}}\right]_{(2)} \left(\text{D}(\hat{\boldsymbol{c}}(\tau)) \hspace{-0.05cm}\otimes\hspace{-0.05cm} \hat{\boldsymbol{H}}^{\text{T}}\right)^{\text{T}}\right)^{\text{T}} \hspace{-0.10cm}\diamond\hspace{-0.05cm} \boldsymbol{I}_{M}\right]^{\dagger} \hspace{-0.15cm} \text{vec}(\left[\hat{\boldsymbol{\mathcal{F}}}\right]_{(2)})
                            \end{align*}
                            \vspace{-0.225cm}
                            \State{Find the \ac{LS} estimate of $\boldsymbol{c}(\tau)$ as}
                            \vspace{-0.225cm}
                            \begin{align*}
                                \hat{\boldsymbol{c}}(\tau) &\hspace{-0.05cm}=\hspace{-0.05cm} \left[\left(\left[\boldsymbol{\mathcal{X}}\right]_{(3)} \left(\text{D}(\hat{\boldsymbol{d}}(\nu)) \hspace{-0.05cm}\otimes\hspace{-0.05cm} \hat{\boldsymbol{H}}^{\text{T}}\right)^{\text{T}}\right)^{\text{T}} \hspace{-0.10cm}\diamond\hspace{-0.05cm} \boldsymbol{I}_{Q}\right]^{\dagger} \hspace{-0.15cm}\text{vec}(\left[\hat{\boldsymbol{\mathcal{F}}}\right]_{(3)})
                            \end{align*}
                            \vspace{-0.225cm}
                            \State{Find the \ac{LS} estimate of $\boldsymbol{H}$ as}
                            \vspace{-0.15cm}
                            \begin{align*}
                                \hat{\boldsymbol{H}} &= \left[\left[\hat{\boldsymbol{\mathcal{F}}}\right]_{(1)} \left(\left[\boldsymbol{\mathcal{X}}\right]_{(1)} \left(\text{D}(\hat{\boldsymbol{c}}(\tau)) \otimes \text{D}(\hat{\boldsymbol{d}}(\nu))\right)^{\text{T}}\right)^{\dagger}\right]^{\text{T}}
                            \end{align*}
                            \vspace{-0.225cm}
                            \State{Update $e(i) = ||\hat{\mathcal{F}} - \hat{\mathcal{F}}(i)||^{2}_{\text{F}}$}
                        \EndWhile
                        \State{\textbf{return} $\hat{\boldsymbol{d}}(\nu)$, $\hat{\boldsymbol{c}}(\tau)$ and $\hat{\boldsymbol{H}}$.}
                        \State{Acquire $\nu$, $\tau$, and $\boldsymbol{\theta}$ by applying $1$D \ac{ESPRIT} on $\hat{\boldsymbol{d}}(\nu)$ and $\hat{\boldsymbol{c}}(\tau)$, and 2D \ac{ESPRIT} on $\left[\hat{\boldsymbol{\mathcal{P}}}\right]_{(3)}$, respectively \cite{tschudin1999comparison,feng2017comparison}.}
                        \State{\textbf{return} $\hat{\nu}$, $\hat{\tau}$, and $\hat{\boldsymbol{\theta}}$.}
                    \end{algorithmic}
                \end{algorithm} 
                \textbf{Second Stage:} By tensorizing $\hat{\boldsymbol{F}}(\tau, \nu) = \left(\boldsymbol{H}^{\text{T}} \boldsymbol{X} \text{D}\left[\boldsymbol{c}(\tau) \otimes \boldsymbol{d}(\nu)\right]\right)^{\text{T}} = \left[\text{D}(\boldsymbol{c}(\tau)) \otimes \text{D}(\boldsymbol{d}(\nu))\right] \boldsymbol{X}^{\text{T}} \boldsymbol{H}$, we obtain a $3$rd order tensor given by
                \begin{align}
                    \hat{\boldsymbol{\mathcal{F}}} &= \boldsymbol{\mathcal{X}} \times_{1} \boldsymbol{H}^{\text{T}} \times_{2} \text{D}(\boldsymbol{d}(\nu)) \times_{3} \text{D}(\boldsymbol{c}(\tau)) \in \mathbb{C}^{N \times M \times Q}, \label{eq:3rd_order_tucker_2}
                \end{align}
                whose core tensor is represented by $\boldsymbol{\mathcal{X}} \in \mathbb{C}^{L \times M \times Q}$. The following matrix unfoldings can be obtained:
                \begin{align}
                    \left[\hat{\boldsymbol{\mathcal{F}}}\right]_{(1)} &\hspace{-0.15cm}=\hspace{-0.05cm} \boldsymbol{H}^{\text{T}} \left[\boldsymbol{\mathcal{X}}\right]_{(1)} \left(\text{D}(\boldsymbol{c}(\tau)) \otimes \text{D}(\boldsymbol{d}(\nu))\right)^{\text{T}} \hspace{-0.05cm}\in\hspace{-0.05cm} \mathbb{C}^{N \times M Q}, \label{eq:stage_2_tucker_1_unfolding_1} \\
                    \left[\hat{\boldsymbol{\mathcal{F}}}\right]_{(2)} &\hspace{-0.15cm}=\hspace{-0.05cm} \text{D}(\boldsymbol{d}(\nu)) \left[\boldsymbol{\mathcal{X}}\right]_{(2)} \left(\text{D}(\boldsymbol{c}(\tau)) \otimes \boldsymbol{H}^{\text{T}}\right)^{\text{T}} \hspace{-0.05cm}\in\hspace{-0.05cm} \mathbb{C}^{M \times N Q}, \label{eq:stage_2_tucker_1_unfolding_2} \\ 
                    \left[\hat{\boldsymbol{\mathcal{F}}}\right]_{(3)} &\hspace{-0.15cm}=\hspace{-0.05cm} \text{D}(\boldsymbol{c}(\tau)) \left[\boldsymbol{\mathcal{X}}\right]_{(3)} \left(\text{D}(\boldsymbol{d}(\nu)) \otimes \boldsymbol{H}^{\text{T}}\right)^{\text{T}} \hspace{-0.05cm}\in\hspace{-0.05cm} \mathbb{C}^{Q \times N M}. \label{eq:stage_2_tucker_1_unfolding_3}
                \end{align}
                
                We now solve the following problem
                \begin{equation}
                    \hspace{-0.3cm} \left\{\hat{\boldsymbol{H}}, \hat{\boldsymbol{d}}(\nu), \hat{\boldsymbol{c}}(\tau)\right\} \hspace{-0.05cm}=\hspace{-0.15cm} \underset{\boldsymbol{H}, \boldsymbol{d}(\nu), \boldsymbol{c}(\tau)}{\text{arg min}} \left|\left|\begin{aligned}\hat{\boldsymbol{\mathcal{F}}} - \boldsymbol{\mathcal{X}} \times_{1} \boldsymbol{H}^{\text{T}} \times_{2}& \\ \text{D}(\boldsymbol{d}(\nu)) \times_{3} \text{D}(\boldsymbol{c}(\tau))&\end{aligned}\right|\right|^{2}_{\text{F}}. \label{eq:tensor_fit_problem_2}
                \end{equation}
                Applying the property $\text{vec}(\boldsymbol{A} \text{D}\left(\boldsymbol{b}\right) \boldsymbol{C}) = (\boldsymbol{C}^{\text{T}} \diamond \boldsymbol{A})\boldsymbol{b}$ to Equations (\ref{eq:stage_2_tucker_1_unfolding_2}) and (\ref{eq:stage_2_tucker_1_unfolding_3}), we have
                \begin{align}
                    \hspace{-0.1cm}\text{vec}(\left[\hat{\boldsymbol{\mathcal{F}}}\right]_{(2)}) &\hspace{-0.05cm}=\hspace{-0.05cm} \left[\hspace{-0.05cm}\left(\left[\boldsymbol{\mathcal{X}}\right]_{(2)} \left(\text{D}(\boldsymbol{c}(\tau)) \hspace{-0.05cm}\otimes\hspace{-0.05cm} \boldsymbol{H}^{\text{T}}\right)^{\text{T}}\right)^{\text{T}} \hspace{-0.15cm}\diamond\hspace{-0.05cm} \boldsymbol{I}_{M}\hspace{-0.05cm} \right] \boldsymbol{d}(\tau), \label{eq:stage_2_tucker_1_unfolding_1_vec} \\
                    \hspace{-0.1cm}\text{vec}(\left[\hat{\boldsymbol{\mathcal{F}}}\right]_{(3)}) &\hspace{-0.05cm}=\hspace{-0.05cm} \left[\hspace{-0.05cm}\left(\left[\boldsymbol{\mathcal{X}}\right]_{(3)} \left(\text{D}(\boldsymbol{d}(\nu)) \hspace{-0.05cm}\otimes\hspace{-0.05cm} \boldsymbol{H}^{\text{T}}\right)^{\text{T}}\right)^{\text{T}} \hspace{-0.15cm}\diamond\hspace{-0.05cm} \boldsymbol{I}_{Q} \hspace{-0.05cm}\right] \boldsymbol{c}(\tau). \label{eq:stage_2_tucker_1_unfolding_2_vec}
                \end{align}
                Similar to the first stage, minimizing \eqref{eq:tensor_fit_problem_2} in the \ac{LS} sense with respect to each factor matrix leads to
                \begin{equation}
                    \hat{\boldsymbol{d}}(\nu) = \underset{\boldsymbol{d}(\nu)}{\text{arg min}} \left|\left|\begin{aligned}\text{vec}(\left[\hat{\boldsymbol{\mathcal{F}}}\right]_{(2)}) - [(\left[\boldsymbol{\mathcal{X}}\right]_{(2)}& \\ \left(\text{D}(\boldsymbol{c}(\tau)) \otimes \boldsymbol{H}^{\text{T}}\right)^{\text{T}})^{\text{T}} \diamond \boldsymbol{I}_{M}] \boldsymbol{d}(\tau)& \end{aligned}\right|\right|^{2}_{\text{F}}, \label{eq:stage_2_tucker_1_ls_2}
                \end{equation} 
                \vspace{-0.1cm}
                \begin{equation}
                    \hat{\boldsymbol{c}}(\tau) = \underset{\boldsymbol{c}(\tau)}{\text{arg min}} \left|\left|\begin{aligned}\text{vec}(\left[\hat{\boldsymbol{\mathcal{F}}}\right]_{(3)}) - [(\left[\boldsymbol{\mathcal{X}}\right]_{(3)}& \\ \left(\text{D}(\boldsymbol{d}(\nu)) \otimes \boldsymbol{H}^{\text{T}}\right)^{\text{T}})^{\text{T}} \diamond \boldsymbol{I}_{Q}] \boldsymbol{c}(\tau)& \end{aligned}\right|\right|^{2}_{\text{F}}, \label{eq:stage_2_tucker_2_ls_3} 
                \end{equation}
                \vspace{-0.1cm}
                \begin{equation}
                    \hat{\boldsymbol{H}} = \underset{\boldsymbol{H}}{\text{arg min}} \left|\left|\begin{aligned}\left[\hat{\boldsymbol{\mathcal{F}}}\right]_{(1)} -& \\  \boldsymbol{H}^{\text{T}} \left[\boldsymbol{\mathcal{X}}\right]_{(1)} \left(\text{D}(\boldsymbol{c}(\tau)) \otimes \text{D}(\boldsymbol{d}(\nu))\right)^{\text{T}}&\end{aligned}\right|\right|^{2}_{\text{F}}, \label{eq:stage_2_tucker_1_ls_1}
                \end{equation} 
                \vspace{-0.2cm}
                whose solutions are given by
                \begin{align} 
                    \hspace{-0.1cm}\hat{\boldsymbol{d}}(\nu) &\hspace{-0.05cm}=\hspace{-0.1cm} \left[\hspace{-0.1cm}\left(\left[\boldsymbol{\mathcal{X}}\right]_{(2)} \hspace{-0.1cm} \left(\text{D}(\boldsymbol{c}(\tau)) \hspace{-0.05cm}\otimes\hspace{-0.05cm} \boldsymbol{H}^{\text{T}}\right)^{\text{T}}\right)^{\text{T}} \hspace{-0.1cm}\diamond \boldsymbol{I}_{M}\right]^{\dagger} \hspace{-0.1cm}\text{vec}(\left[\hat{\boldsymbol{\mathcal{F}}}\right]_{(2)}), \label{eq:2nd_stage_pseudoinverse_1} \\
                    \hspace{-0.1cm}\hat{\boldsymbol{c}}(\tau) &\hspace{-0.05cm}=\hspace{-0.1cm} \left[\hspace{-0.1cm}\left(\left[\boldsymbol{\mathcal{X}}\right]_{(3)} \hspace{-0.1cm} \left(\text{D}(\boldsymbol{d}(\nu)) \hspace{-0.05cm}\otimes\hspace{-0.05cm} \boldsymbol{H}^{\text{T}}\right)^{\text{T}}\right)^{\text{T}} \hspace{-0.1cm}\diamond \boldsymbol{I}_{Q}\right]^{\dagger} \hspace{-0.1cm}\text{vec}(\left[\hat{\boldsymbol{\mathcal{F}}}\right]_{(3)}), \label{eq:2nd_stage_pseudoinverse_2} \\
                    \hat{\boldsymbol{H}} &= \left[\left[\hat{\boldsymbol{\mathcal{F}}}\right]_{(1)} \left(\left[\boldsymbol{\mathcal{X}}\right]_{(1)} \left(\text{D}(\boldsymbol{c}(\tau)) \otimes \text{D}(\boldsymbol{d}(\nu))\right)^{\text{T}}\right)^{\dagger}\right]^{\text{T}}. \label{eq:2nd_stage_pseudoinverse_3}
                \end{align}
                
            Using the estimates of $\hat{\boldsymbol{d}}(\nu)$, $\hat{\boldsymbol{c}}(\tau)$, and $\left[\hat{\boldsymbol{\mathcal{P}}}\right]_{(3)}$, we respectively extract the target parameters $\hat{\nu}$, $\hat{\tau}$, and $\hat{\boldsymbol{\theta}}$ using a high-resolution estimation algorithm, e.g., \ac{ESPRIT} \cite{tschudin1999comparison,feng2017comparison}. The proposed method is outlined in Algorithm \ref{alg:proposed}. To determine the computational complexity of the proposed approach, note that computing the pseudoinverse of a matrix $\mathbf{A} \in \mathbb{C}^{I \times J}$, with $I > J$, and its rank-$R$ SVD approximation have complexities $\mathcal{O}(I J^{2})$ and $\mathcal{O}(I J R)$, respectively. The complexity required to complete the $1$st and $2$dn stage is respectively of order $\mathcal{O}(\text{\acs{ALS}}_{\text{iter}1}(N^2 K [M Q (1 + L N^2) + L]))$ and $\mathcal{O}(\text{\acs{ALS}}_{\text{iter}2}(N [M Q(M^2 + Q^2 ) + L^2]))$, where $\text{\acs{ALS}}_{\text{iter}1}$ and $\text{\acs{ALS}}_{\text{iter}2}$ are the number of iterations required in each stage.
                
        \subsection{Identifiability and Uniqueness} \label{sec:identifiability_and_uniqueness}
            To ensure uniqueness when solving (\ref{eq:3rd_order_tucker}) and (\ref{eq:3rd_order_tucker_2}), knowledge of the \acs{RIS} phase-shift matrix, $\boldsymbol{W}$, and the transmitted symbols, $\boldsymbol{X}$, are assumed to be available at the \ac{BS}, which is a reasonable assumption. Given knowledge of both $\boldsymbol{W}$ and $\boldsymbol{X}$, the scaling ambiguities associated with the first stage are given by 
            $\hspace{-0.1cm} \boldsymbol{H} \hspace{-0.1cm}=\hspace{-0.1cm} \hat{\boldsymbol{H}} \boldsymbol{\Lambda}_{h}, \boldsymbol{F}(\tau,\nu) \hspace{-0.1cm}=\hspace{-0.1cm} \hat{\boldsymbol{F}}(\tau,\nu) \boldsymbol{\Lambda}_{f}, \text{ and } \left[\boldsymbol{\mathcal{P}}\right]_{(3)} \hspace{-0.1cm}=\hspace{-0.1cm} \boldsymbol{\Lambda}^{-1}_{h} [\hat{\boldsymbol{\mathcal{P}}}]_{(3)} \boldsymbol{\Lambda}^{-1}_{f}$, where $\boldsymbol{\Lambda}_{h}$ and $\boldsymbol{\Lambda}_{f}$ are arbitrary diagonal matrices. The scaling ambiguities associated with the second stage are given by $\boldsymbol{d}(\nu) = \lambda_{\nu} \hat{\boldsymbol{d}}(\nu), \boldsymbol{c}(\tau) = \lambda_{\tau} \hat{\boldsymbol{c}}(\tau), \text{ and } \boldsymbol{H} = \hat{\boldsymbol{H'}} \boldsymbol{\Lambda}_{h'}$, 
            where $\boldsymbol{\Lambda}_{h} = \text{D}\left(\boldsymbol{H}_{(1,:)} \oslash \hat{\boldsymbol{H}}_{(1,:)}\right)$, $\boldsymbol{\Lambda}_{h'} = \text{D}\left(\boldsymbol{H}_{(1,:)} \oslash \hat{\boldsymbol{H'}}_{(1,:)}\right)$, $\boldsymbol{\Lambda}_{f} = \text{D}\left(\boldsymbol{F}(\tau,\nu)_{(1,:)} \oslash \hat{\boldsymbol{F}}(\tau,\nu)_{(1,:)}\right)$, $\lambda_{\nu} = [\boldsymbol{d}(\nu)]_{1} \oslash [\hat{\boldsymbol{d}}(\nu)]_{1}$, and $\lambda_{\tau} = [\boldsymbol{c}(\tau)]_{1} \oslash [\hat{\boldsymbol{c}}(\tau)]_{1}$ \cite{sokal2019semi}. To ensure the uniqueness of the estimates in our tensor model, it is required that $K \geq N^2$, and $M Q \geq L$ so that the pseudoinverses (\ref{eq:1st_stage_pseudoinverse_1})-(\ref{eq:1st_stage_pseudoinverse_3}) and (\ref{eq:2nd_stage_pseudoinverse_1})-(\ref{eq:2nd_stage_pseudoinverse_3}) are unique.
  
    \section{Simulation Results}
        In this section, we evaluate the performance of the proposed \ac{NTBS} algorithm. We design the \acs{RIS} phase-shift matrix $\boldsymbol{W}$ as a truncated discrete Fourier transform matrix of size $N$ with $K$ columns. The \acs{RIS} elevation and azimuth angles of arrival and departure are randomly generated from a uniform distribution between $[0, \pi/2]$. The parameter estimation accuracy is evaluated in terms of the RMSE defined as $\text{\ac{RMSE}}(x) = \sqrt{\mathbb{E}\left[\left|x^{(v)} - \hat{x}^{(v)}\right|^{2}/\left|x^{(v)}\right|^{2}\right]}$, where $x^{(v)}$ is one of the target parameters estimated in the $v$th experiment, and $V = 5 \times 10^{3}$ is the number of independent channel realizations. The \ac{SNR} is defined as $\text{SNR} = ||\ten{Y}||^{2}_{\text{F}}\sigma^{2}_{\ten{Z}}/||\ten{Z}||^{2}_{\text{F}}$ where $\sigma^{2}_{\ten{Z}}$ is the noise variance. Table \ref{tab:parameters} summarizes the parameters of the simulation scenario. 
        
\begin{figure}[!t]
            \centering
            \begin{minipage}{.485\columnwidth}
                \includegraphics[width=.975\linewidth]{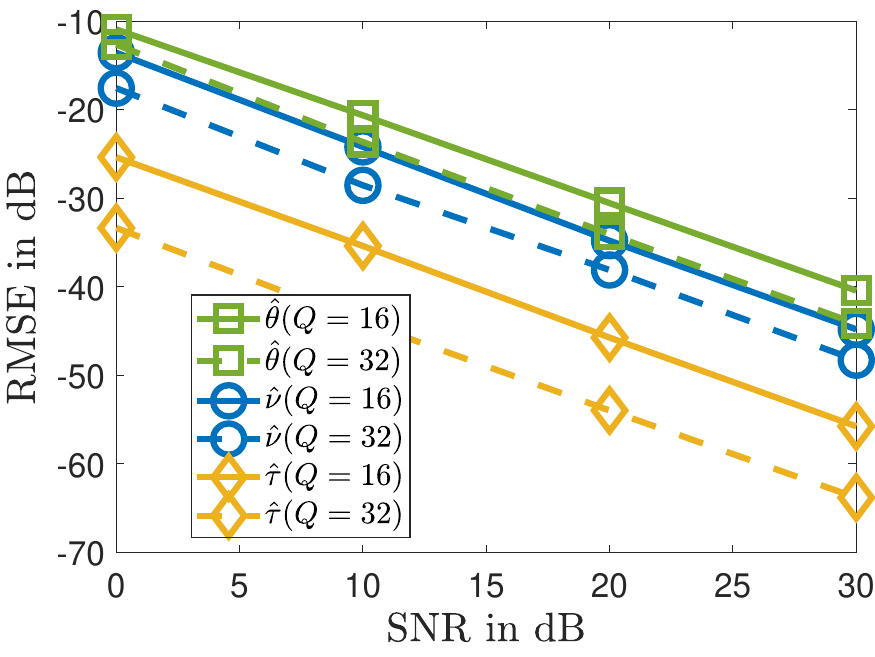}
                \caption{\ac{RMSE} as a function of the number of subcarriers.}
                \label{fig:rmse_case_1}
            \end{minipage}
            \begin{minipage}{.485\columnwidth}
                \includegraphics[width=.975\linewidth]{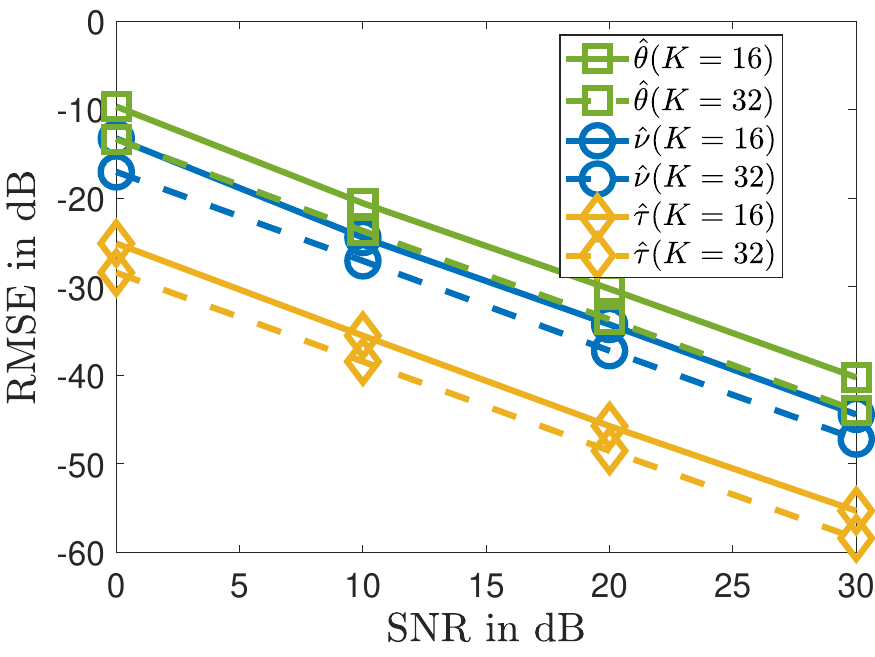}
                \caption{\ac{RMSE} as a function of the number of blocks.}
                \label{fig:rmse_case_2}
            \end{minipage}
        \end{figure}

\begin{figure}[!t]
            \centering
            \begin{minipage}{.485\columnwidth}
                \includegraphics[width=.975\linewidth]{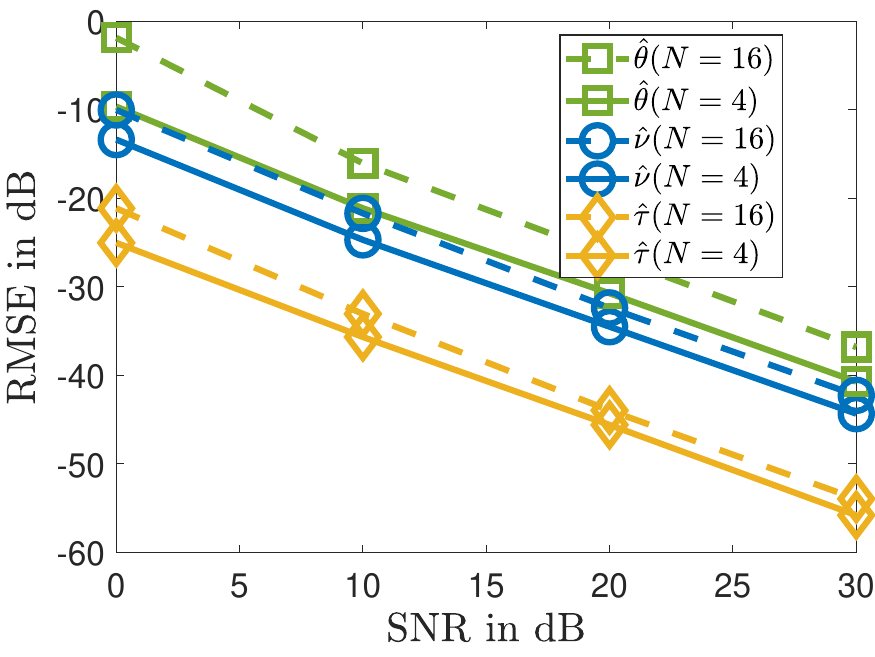}
                \caption{\ac{RMSE} as a function of the number of RIS elements.}
                \label{fig:rmse_case_3}
            \end{minipage}
            \begin{minipage}{.485\columnwidth}
                \includegraphics[width=.975\linewidth]{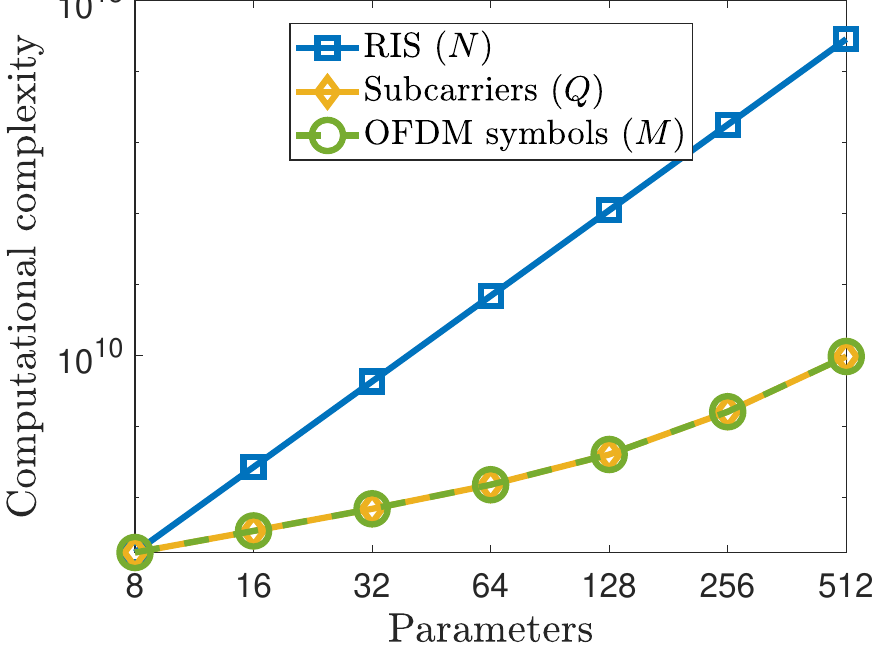}
                \caption{Computational complexity of Algorithm \ref{alg:proposed}.}
                \label{fig:computational_complexity_case_1}
            \end{minipage}
        \end{figure}

        Figure \ref{fig:rmse_case_1} evaluates the \ac{RMSE} performance of the proposed \ac{NTBS} algorithm in terms of as the number of subcarriers $Q$ varies. The simulations show that increasing $Q$ improves the estimation performance for all parameters. Notably, the performance enhancement is more significant for delay estimation, as we achieve a higher resolution of the dimension linked to the delay in our tensor models in Equation (\ref{eq:3rd_order_tucker}). Likewise, in Figure \ref{fig:rmse_case_2}, we evaluate the performance as a function of the number of blocks, $K$, composing the sensing protocol in Figure \ref{fig:transmission_protocol}. We observe behavior similar to Figure \ref{fig:rmse_case_1}, where increasing the number of blocks leads to an enhancement in performance mainly due to an increase in the resolution of the tensor model in Equation (\ref{eq:3rd_order_tucker}). In Figure \ref{fig:rmse_case_3}, we evaluate the impact of the number of \acs{RIS} reflecting elements on performance. Increasing \(N\) leads to a deterioration in the \ac{RMSE} since adding more parameters to estimate does not improve resolution in the first stage of Equation (\ref{eq:3rd_order_tucker}), leading to propagated estimation errors in the second stage. 
        
        Figure \ref{fig:computational_complexity_case_1} illustrates the computational complexity of the proposed \ac{NTBS} algorithm for three different scenarios. The parameters in each scenario are set according to Table \ref{tab:parameters}, with only one parameter varying at a time. We see that the computational complexity increases linearly with the number of \acs{RIS} elements. In contrast, variations in the number of subcarriers and \ac{OFDM} symbols have a minor impact on complexity. In addition to the \ac{RMSE} performance results, the complexity curve demonstrates that the proposed algorithm benefits from an increase in $Q$, without a significant rise in the complexity required to solve Algorithm \ref{alg:proposed}.

        \begin{table}[!t]
            \centering
            \caption{Simulation parameters.}
            \label{tab:parameters}
            \resizebox{0.65\columnwidth}{!}{
            \begin{tabular}{|c|c|}
                \hline
                \acs{RIS} elements & $N_{x} N_{y} = 4 \times 4 = 16$ \\ \hline
                \acs{RIS} spacing & $d_{x} = d_{y} = \lambda/2$ \\ \hline
                \ac{AoA} generation & $\{\phi_{\text{az}}, \phi_{\text{el}}\} \sim \text{U}(0, 90^{\circ})$ \\ \hline
                \ac{AoD} generation & $\{\theta_{\text{az}}, \theta_{\text{el}}\} \sim \text{U}(0, 90^{\circ})$ \\ \hline
                Wavelength & $1.07 \times 10^{-2}$ m \\ \hline
                Carrier frequency & $28$ GHz \\ \hline
                Subcarrier spacing $\Delta f$ & $120$ KHz \\ \hline
                Symbol duration & $1/\Delta f$ \\ \hline
                Number of symbols & $64$ \\ \hline
                Number of subcarriers & $16$ \\ \hline
                Distance \ac{BS} - \acs{RIS} & $10$ m \\ \hline
                Distance \acs{RIS} - target & $5$ m \\ \hline
                RCS & $2 \text{m}^{2}$ \\ \hline
            \end{tabular}
            }
        \end{table}

    \section{Conclusion}
        We have proposed an \acs{NTBS} model that fully exploits the intrinsic multidimensional nature of the received echo signal at the \ac{BS} in a monostatic radar system, where the goal is to estimate the parameters of a single target. Our proposed \ac{NTBS} algorithm estimates the delay, Doppler, and angle of the target using a two-stage \ac{ALS} + \ac{ESPRIT} approach. Our simulation results study the impact of the system parameters on the \ac{RMSE} performance and complexity. Future work includes an extension of the \ac{NTBS} algorithm to the multi-target scenario.
        
    \bibliographystyle{IEEEtran}
    \bibliography{references}

% Generated by IEEEtran.bst, version: 1.14 (2015/08/26)
\begin{thebibliography}{10}
\providecommand{\url}[1]{#1}
\csname url@samestyle\endcsname
\providecommand{\newblock}{\relax}
\providecommand{\bibinfo}[2]{#2}
\providecommand{\BIBentrySTDinterwordspacing}{\spaceskip=0pt\relax}
\providecommand{\BIBentryALTinterwordstretchfactor}{4}
\providecommand{\BIBentryALTinterwordspacing}{\spaceskip=\fontdimen2\font plus
\BIBentryALTinterwordstretchfactor\fontdimen3\font minus \fontdimen4\font\relax}
\providecommand{\BIBforeignlanguage}[2]{{%
\expandafter\ifx\csname l@#1\endcsname\relax
\typeout{** WARNING: IEEEtran.bst: No hyphenation pattern has been}%
\typeout{** loaded for the language `#1'. Using the pattern for}%
\typeout{** the default language instead.}%
\else
\language=\csname l@#1\endcsname
\fi
#2}}
\providecommand{\BIBdecl}{\relax}
\BIBdecl

\bibitem{Rui_Zhang_2021}
Q.~Wu, S.~Zhang, B.~Zheng, C.~You, and R.~Zhang, ``Intelligent reflecting surface-aided wireless communications: A tutorial,'' \emph{IEEE Transactions on Communications}, vol.~69, no.~5, pp. 3313--3351, 2021.

\bibitem{Fan_2022}
F.~Liu, Y.~Cui, C.~Masouros, J.~Xu, T.~X. Han, Y.~C. Eldar, and S.~Buzzi, ``Integrated sensing and communications: Toward dual-functional wireless networks for {6G} and beyond,'' \emph{IEEE Journal on Selected Areas in Communications}, vol.~40, no.~6, pp. 1728--1767, 2022.

\bibitem{zhang2022metaradar}
H.~Zhang, H.~Zhang, B.~Di, K.~Bian, Z.~Han, and L.~Song, ``{MetaRadar}: Multi-target detection for reconfigurable intelligent surface aided radar systems,'' \emph{IEEE Trans. Wireless Commun.}, vol.~21, no.~9, pp. 6994--7010, 2022.

\bibitem{rihan2022spatial}
M.~Rihan, E.~Grossi, L.~Venturino, and S.~Buzzi, ``Spatial diversity in radar detection via active reconfigurable intelligent surfaces,'' \emph{IEEE Signal Processing Letters}, vol.~29, pp. 1242--1246, 2022.

\bibitem{song2023intelligent}
X.~Song, J.~Xu, F.~Liu, T.~X. Han, and Y.~C. Eldar, ``Intelligent reflecting surface enabled sensing: Cram{\'e}r-{R}ao bound optimization,'' \emph{IEEE Transactions on Signal Processing}, 2023.

\bibitem{shao2024target}
X.~Shao and R.~Zhang, ``Target-mounted intelligent reflecting surface for secure wireless sensing,'' \emph{IEEE Trans. Wireless Commun.}, 2024.

\bibitem{kemal2024ris}
M.~Kemal~Ercan, M.~Furkan~Keskin, S.~Gezici, and H.~Wymeersch, ``{RIS}-aided {NLoS} monostatic sensing under mobility and angle-doppler coupling,'' \emph{arXiv e-prints}, pp. arXiv--2401, 2024.

\bibitem{10639156}
K.~B.~A. Benício, B.~Sokal, A.~L.~F. de~Almeida, Fazal-E-Asim, B.~Makki, and G.~Fodor, ``Low-complexity tensor-based monostatic sensing for {IRS}-assisted communication systems,'' in \emph{Proc. 19th International Symposium on Wireless Communication Systems (ISWCS)}, 2024.

\bibitem{ximenes2015semi}
L.~R. Ximenes, G.~Favier, and A.~L. de~Almeida, ``Semi-blind receivers for non-regenerative cooperative mimo communications based on nested parafac modeling,'' \emph{IEEE transactions on signal processing}, vol.~63, no.~18, pp. 4985--4998, 2015.

\bibitem{tschudin1999comparison}
M.~Tschudin, C.~Brunner, T.~Kurpjuhn, M.~Haardt, and J.~A. Nossek, ``Comparison between unitary {ESPRIT} and {SAGE} for {3-D} channel sounding,'' in \emph{Proc. 49th IEEE VTC'99}, vol.~2, 1999, pp. 1324--1329.

\bibitem{feng2017comparison}
R.~Feng, Y.~Liu, J.~Huang, J.~Sun, and C.-X. Wang, ``Comparison of {MUSIC}, unitary {ESPRIT}, and {SAGE} algorithms for estimating {3D} angles in wireless channels,'' in \emph{Proc. IEEE/CIC International Conference on Communications in China (ICCC)}, 2017.

\bibitem{Asim_2021}
Fazal-E-Asim, F.~Antreich, C.~C. Cavalcante, A.~L.~F. de~Almeida, and J.~A. Nossek, ``Two-dimensional channel parameter estimation for millimeter-wave systems using {Butler} matrices,'' \emph{IEEE Trans. Wireless Commun.}, vol.~20, no.~4, pp. 2670--2684, 2021.

\bibitem{comon2009tensor}
P.~Comon, X.~Luciani, and A.~L.~F. de~Almeida, ``Tensor decompositions, alternating least squares and other tales,'' \emph{Journal of Chemometrics}, vol.~23, no. 7-8, pp. 393--405, 2009.

\bibitem{de2021channel}
G.~T. de~Ara{\'u}jo, A.~L. de~Almeida, and R.~Boyer, ``Channel estimation for intelligent reflecting surface assisted mimo systems: A tensor modeling approach,'' \emph{IEEE Journal of Selected Topics in Signal Processing}, vol.~15, no.~3, pp. 789--802, 2021.

\bibitem{benicio2023tensor}
K.~Ben{\'\i}cio, A.~L. de~Almeida, B.~Sokal, B.~Makki, G.~Fodor \emph{et~al.}, ``Tensor-based modeling/estimation of static channels in {IRS}-assisted {MIMO} systems,'' \emph{XLI Brazilian Symposium on Telecommunications and Signal Processing (SBrT)}, 2023.

\bibitem{benicio2023tensor_wcl}
K.~B. Ben{\'\i}cio, A.~L. de~Almeida, B.~Sokal, B.~Makki, G.~Fodor \emph{et~al.}, ``Tensor-based channel estimation and data-aided tracking in {IRS}-assisted {MIMO} systems,'' \emph{IEEE Wireless Comms. Letters}, 2023.

\bibitem{sokal2019semi}
B.~Sokal, M.~Haardt, and A.~L.~F. de~Almeida, ``Semi-blind receiver for two-hop {MIMO} relaying systems via selective {Kronecker} product modeling,'' in \emph{Proc. 8th IEEE CAMSAP'19}, 2019, pp. 714--718.

\end{thebibliography}
\end{document}